\documentclass[preprint2]{aastex}

\slugcomment{To appear in the Astrophysical Journal}

\shorttitle{ A scaling law for interstellar depletions }
\shortauthors{G. Vladilo}

\begin{document}
   \title{ A  scaling law for 
interstellar depletions }

\author{ Giovanni  Vladilo } 
\affil{Osservatorio Astronomico di Trieste,
              Via G.B. Tiepolo 11, I-34131 Trieste}
\email{vladilo@ts.astro.it}


\begin{abstract}
An analytical expression is presented that allows  
gas-to-dust
elemental depletions  to be estimated
in interstellar environments of different types,
including Damped Ly\,$\alpha$ systems,
by scaling  an arbitrary depletion pattern chosen
as a reference.
As an improvement on previous work,
 the scaling relation allows the  dust chemical composition to vary
   and includes a set of parameters which describe how
sensitive  the dust composition is to changes in both
 the dust-to-metals ratio
and   the  composition of the medium.
These parameters can be estimated empirically from studies of Galactic
and extragalactic depletion patterns. 
The  scaling law is able to fit all the typical
depletion patterns   of 
 the   Milky Way ISM
 (\emph{cold disk}, \emph{warm disk}, 
and \emph{warm halo})
with a single set of  parameters, by only varying the dust-to-metals
ratio. 
The dependence of the scaling law on the abundances
of the medium has been tested using interstellar
observations of the Small Magellanic Cloud (SMC),
for which peculiar depletion patterns have been
reported in  literature.   
The scaling law is able to fit these  depletion patterns
assuming that the   SMC relative abundances are slightly non solar. 
\end{abstract}
   

\section{Introduction} 

It is well known that elemental abundances 
measured in the  interstellar gas of the solar vicinity
are generally depleted with respect to the
solar values (Morton 1974; Jenkins, Savage \& Spitzer 1986).  
The commonly accepted interpretation of this depletion effect
is  that a fraction of these elements is not detected in the gas  
because it is locked into dust grains.  
Elemental  depletions are estimated by comparing the abundances
measured in the gas with the abundances of the medium (gas plus dust).
Intrinsic solar abundances are usually
adopted in local interstellar studies, even though the interstellar abundance standard is not yet
firmly established   (Savage \& Sembach 1996, hereafter SS96; Sofia \& Meyer 2001).
Different elements may have very different values of depletion,
the so called "refractory" elements being almost completely
 in  dust form.  
  In addition,
 elemental depletions can  vary significantly among
different lines of sight.  
In spite of this complex behavior, 
interstellar regions with similar physical conditions are characterized by
similar depletions (Spitzer 1985; Jenkins et al. 1986). 
In the Galactic disk the  cold  gas has higher depletions than
the warm gas;  in the halo the warm gas  shows      even lower 
depletions (SS96).  

Elemental  depletions  constitute an important piece of information for
solving the complex puzzle of
the origin and evolution of interstellar dust 
(Tielens \& Allamandola 1987 and refs. therein).  
Studies of depletions in the Galaxy
and in the Magellanic Clouds (MCs)
have received new  impulse 
(Roth \& Blades 1995; Welty et al. 1997, 2001)  
owing to their importance in the 
interpretation of abundance patterns measured in 
the QSO absorbers  of highest \ion{H}{1} column density, namely the 
Damped Ly $\alpha$ systems (DLAs). 
These systems  originate in
intervening galaxies or protogalaxies and 
their abundances yield unique information on  galactic
chemical evolution at different redshifts
(Lu et al. 1996; Prochaska \& Wolfe 1999; Molaro et al. 2000).  
The evidence that dust depletion may significantly affect the
  abundances of DLAs is quite compelling 
(Pettini et al. 1994; Lauroesch et al. 1996; 
Hou et al. 2001; Prochaska \& Wolfe 2002). 
In order to cope with the problem of   depletion in DLAs
some authors have  concentrated    
on studies of  non-refractory elements
(Pettini et al. 1997;  
Centuri\`on et al. 2000; 
Vladilo et al. 2000)
 and of   systems with small dust content (Pettini et al. 2000, Molaro et al. 2000). 
At the same time, 
  formalisms have been developed to quantify
the effect of dust depletions on DLA abundances 
(Kulkarni et al. 1997; Vladilo 1998; 
Savaglio, Panagia \& Stiavelli 2000). 
These formalisms assume that  the dust in DLAs is similar 
to Galactic interstellar dust.  
Depletions in  DLAs have been modeled  scaling Galactic depletions   
by  allowing    variations of the dust-to-metals
ratios, but not of the dust chemical composition.  In addition,
the fact that the dust composition may be affected by variations
of  the intrinsic abundances of the intervening galaxies
has not been considered. 

In this work a scaling law
of interstellar depletions is derived 
that allows  the dust chemical composition to vary 
 according to
changes in the dust-to-metals ratio and/or to changes
in the overall abundances of the medium (Section 2). 
Rather than assuming {\it ad hoc} types of dust, a set of
parameters is introduced to  describe how sensitive the dust composition is 
to changes in the physical and chemical properties
of the medium. 
In Sections 3 and 4 it is shown that  
such parameters can be estimated
from Galactic and extragalactic interstellar 
data. The conclusions are summarized in Section 5.

\section{A scaling law for interstellar depletions}

The interstellar depletion of an element X is usually defined as  
\begin{equation}
\label{Depletion1}
D_\mathrm{x}= 
\log  \left( { N_\mathrm{x} \over N_\mathrm{H} } \right)_\mathrm{g}  - 
\log \left( {  {\rm X} \over {\rm H}  } \right)_{\rm m} ,
\end{equation}
where 
 $N_\mathrm{x}$ and
$N_\mathrm{H}$ 
are the   column   densities measured  in the gas (Spitzer 1978)  
and
 $( { {\rm X}  \over {\rm H} }  )_{\rm m} $
is the abundance in the medium (gas plus dust). 
In Galactic  studies one usually takes
$( { {\rm X}  \over {\rm H} }  )_{\rm m} = ( { {\rm X}  \over {\rm H} }  )_{\sun} $ 
(see, however,  Sofia \& Meyer 2001).  
For galaxies with non-solar
composition one should instead use the corresponding 
abundance pattern. 
 %
%
%
%
The depletion $D_\mathrm{x}$ does not scale linearly with the quantity
of dust or with its chemical composition. In order to
 derive a scaling law 
  we define instead the \emph{fraction in dust}  ,  
\begin{equation}
\label{fraction0}
f_\mathrm{x}= { N_\mathrm{x,d} \over N_\mathrm{x,m} }  ~,
\end{equation}
where 
 $N_\mathrm{x,d}$ is the column density of atoms X in  the dust and
$N_\mathrm{x,m}$ 
  the   column   density in the medium  (gas plus dust).\footnote{
We consider quantities  integrated
along the line of sight, such as the column density, for consistency
with the empirical definition (\ref{Depletion1}).
An identical treatment could be done using  local quantities instead,
such as volume densities. 
} 
The fraction in dust is simply related to the measured depletion: 
\begin{equation}
\label{fraction2}
f_\mathrm{x} = 1 - 10^{D_\mathrm{x}} ~~. 
\end{equation}
The fraction  $f_\mathrm{x}$ must
scale  linearly with the amount of dust and 
with the abundance of X in dust. 
To derive this relation we express all abundances by number
and relative to the abundance of an arbitrary element Y (e.g. iron).    
We therefore define the \emph{abundance of X in the medium}  
\begin{equation}
\label{RelativeAbundance}
a_\mathrm{x}  = { N_\mathrm{x,m}   \over N_\mathrm{y,m}  } ~,
\end{equation}
and the 
\emph{abundance of X in the dust}  
\begin{equation}
\label{pX}
p_\mathrm{x} = { N_{\rm x,d} \over N_\mathrm{y,d} } ~.
\end{equation}

We also use the element Y as a reference for estimating
the dust content of the system. 
For a given  chemical composition of the dust  
(of the medium)
the total number of atoms of metals in the dust 
(in the medium)  
will be proportional to  $N_\mathrm{y,d}$
(to  $N_\mathrm{y,m}$).
Therefore
we  define the \emph{dust-to-metals ratio} as
\begin{equation}
\label{DustToMetal}
r   =  { N_\mathrm{y,d} \over N_\mathrm{y,m} } ~. 
\end{equation}
This definition is free from any assumption on the extinction properties
of the dust. 
With the above definitions 
the fraction in dust  becomes
\begin{equation}
\label{Fraction}
f_\mathrm{x} = r  ~ a_\mathrm{x}^{-1} ~ p_\mathrm{x}   ~.
\end{equation}
 In previous work  $f_\mathrm{x}$
 was assumed to
scale  linearly with $r$. This is correct
only if the $p_\mathrm{x}$ parameters are constant, i.e. if the
dust composition does not vary when $r$ varies.  
Here the more general situation is considered in which  the abundances
in the dust 
may depend on $r$ and also on $a_\mathrm{x}$, i.e. 
\begin{equation}
\label{Dependence0}
p_\mathrm{x} = p_\mathrm{x}(r,a_\mathrm{x}) ~~.
\end{equation}
Studies of the  Galactic ISM   indicate
that the dust-to-metals ratio is  linked to the 
physical state of the medium. In fact
clouds with similar physical properties have similar levels
of depletion and depletions appear to be lower when the  physical conditions 
become harsher (e.g. higher temperature or kinetic energy) (SS96). 
%
%
It is reasonable to expect that once the physical conditions of the medium are
specified,  also $r$ and
the dust composition characteristic of those particular physical
conditions are determined.  
If we consider $r$ as an indirect indicator of the physical state, then
relation (\ref{Dependence0}) is equivalent to assuming that
\emph{the dust composition is  determined by 
the physical state and by the chemical composition of the medium}.
This assumption is much more realistic than assuming that the
dust composition is constant.  

Therefore, we consider  $r$ and  $a_\mathrm{x}$ as independent variables of our
problem. 
By logarithmic  differentiating Eq. (\ref{Fraction}) 
taking into account the functional dependence
(\ref{Dependence0}), we derive 
\begin{equation}
\label{PercentVariation}
{ d f_\mathrm{x} \over f_\mathrm{x} }
= 
(1+\eta_\mathrm{x}) ~ { d r \over r } +
(\varepsilon_\mathrm{x} -1) ~ {d a_\mathrm{x} \over  a_\mathrm{x} }
~,
\end{equation}
where
\begin{equation}
\label{etax}
 \eta_\mathrm{x} \equiv  {r \over p_\mathrm{x}} ~
  { \partial p_\mathrm{x}  \over \partial r }
\end{equation}
and 
\begin{equation}
\label{epsilonx}
\varepsilon_\mathrm{x} \equiv  { a_\mathrm{x} \over p_\mathrm{x}} ~
 { \partial p_\mathrm{x}  \over \partial a_\mathrm{x} } ~.
\end{equation}
%


We  now consider an interstellar medium  $i$ (with
  $r=r_i$ and $a_\mathrm{x}=a_{\mathrm{x},i}$) 
that undergoes   small variations $\delta r$
and $\delta a_\mathrm{x}$ and is transformed, as a result,
into a medium   $j$ (with
  $r=r_j$ and $a_\mathrm{x}=a_{\mathrm{x},j}$). 
If $\eta_\mathrm{x}$ and $\varepsilon_\mathrm{x}$ 
can be considered constant during such transformation  
we can integrate Eq. (\ref{PercentVariation})
and obtain the scaling law
\begin{equation}
\label{ScalingLawGeneral}
  { f_{\mathrm{x},j} \over f_{\mathrm{x},i} }  =  
\left( { r_{j} \over r_{i} }\right)^{(1+\eta_\mathrm{x})} ~
\left( { a_{\mathrm{x},j} \over a_{\mathrm{x},i} } \right)^{(\varepsilon_\mathrm{x} -1)} ~ .
\end{equation}

The scaling law (\ref{ScalingLawGeneral}), together with Eq. (\ref{Depletion1}), 
relates variations of depletion patterns to  variations of dust-to-metals ratios
and  of elemental abundances. 
This relation has been derived mathematically,
having made  no assumption 
as to the  mechanisms of dust formation, accretion and destruction.
The relation  holds always true for infinitesimal changes  
$\delta r$ and $\delta a_\mathrm{x}$. 
It can be applied to   transformations involving discrete changes
$\Delta r$ and $\Delta a_\mathrm{x}$ as long as 
$\eta_\mathrm{x}$ and $\varepsilon_\mathrm{x}$ 
are constant during the integration of Eq. (\ref{PercentVariation}). 
Thanks to the fact that abundances in the dust
$p_\mathrm{x}$ do not appear explicitly
in the scaling law, it is not necessary to make {\it ad hoc} assumptions
on the dust chemical composition. 
Instead, we must determine the parameters  $\eta_\mathrm{x}$ and
$\varepsilon_\mathrm{x}$. In the following sections we discuss
the physical meaning of such parameters and  show how they
can be empirically determined. 


\subsection{Physical meaning of the parameters $\eta_\mathrm{x}$ and
$\varepsilon_\mathrm{x}$}

The physical meaning of  $\eta_\mathrm{x}$ and $\varepsilon_\mathrm{x}$ 
can be understood by writing
the   definitions (\ref{etax}) and (\ref{epsilonx})
in the following form  
\begin{equation}
\label{etax2}
{\delta p_\mathrm{x} \over p_\mathrm{x} }
 \simeq \eta_\mathrm{x} { \delta r \over r } ~~,
\end{equation}
and 
\begin{equation}
\label{epsilonx2}
{\delta p_\mathrm{x} \over p_\mathrm{x} }
 \simeq \varepsilon_\mathrm{x} { \delta a_\mathrm{x} \over a_\mathrm{x} } ~~. 
\end{equation}

Both parameters describe how sensitive  the dust relative abundances
$p_\mathrm{x}$  are
to changes in the conditions of the medium. 
The parameter $\eta_\mathrm{x}$ indicates what percent variation
of the relative abundance in the dust  occurs in correspondence to a  
percent variation of the dust-to-metals ratio.
Since $r$ is related to the physical conditions of the medium,  
in practice 
$\eta_\mathrm{x}$ indicates how efficiently the abundance in the dust
  is affected by   changes of the physical conditions.  
If $\eta_\mathrm{x}=0$, the relative abundance in the dust of 
the element X  is not affected by
variations of $r$. 

  The parameter $\varepsilon_\mathrm{x}$ indicates the percent variation
of the relative abundance in the dust that occurs in correspondence to a  
percent variation of the relative abundance in the medium.
  In practice,
the parameters $\varepsilon_\mathrm{x}$ indicate  the efficiency of
the mechanisms that   convert
 a   change in the  composition of the medium into a 
 change in the composition of  the dust.
If $\varepsilon_\mathrm{x}=0$, the relative
  abundance of X in the dust is not affected by changes
in the relative abundance in the medium.

The parameters $\eta$ and $\varepsilon$ have special values,
or do not need to be determined,  for the reference element Y.  
In fact, it is easy to see
from (\ref{ScalingLawGeneral}) 
that     $\eta_\mathrm{y}=0$ since, by definition, 
$f_{\mathrm{y},j} / f_{\mathrm{y},i} = r_j/r_i$. On the other hand, because
$a_{\mathrm{y},j}=a_{\mathrm{y},i}=1$,      $\varepsilon_\mathrm{y}$
does not need to be determined to use relation (\ref{ScalingLawGeneral}).  
In the  next two   sections  we show how  it is possible to empirically
determine  the  
parameters $\eta_\mathrm{x}$ and  $\varepsilon_\mathrm{x}$
for the generic element X.

\section{Estimate of the $\eta_\mathrm{x}$ parameters from Galactic studies}

We assume that Galactic interstellar abundances are constant, at least
in the solar vicinity. This implies that $a_{\mathrm{x},j}=a_{\mathrm{x},i}$ and,
as a consequence, 
the scaling law (\ref{ScalingLawGeneral}) becomes
\begin{equation}
\label{ScalingLawISM}
f_{\mathrm{x},j}   =   
\left( {r_j \over r_i} \right)^{\left( 1+\eta_\mathrm{x} \right)} ~
f_{\mathrm{x},i}   ~.
\end{equation}
Given the simplified form of the scaling relation,
the   Galactic ISM is the ideal place to determine the parameters
$\eta_\mathrm{x}$ and to probe whether they are constant or not
in different environments. 
The  fractions in dust $f_\mathrm{x}$ 
and $r=f_\mathrm{y}$
can be determined from the measured
depletions by 
using Eqs.(\ref{fraction2}) and (\ref{Depletion1}). 
By comparing the fractions in dust in two different
types of clouds $i$ and $j$, each type characterized by a specific depletion pattern,
it is possible to derive  
$\eta_{\mathrm{x},ij} = \log (f_{\mathrm{x},j}/f_{\mathrm{x},i} )
/ \log (r_j / r_i) -1 $. 
If  the parameter $\eta_\mathrm{x}$ is constant
for any variation of the dust-to-metals ratio, 
then  a single value
of  $\eta_\mathrm{x}$ should be obtained  from any   pair ($ij$)
of depletion patterns considered. 

\begin{figure}   
\plotone{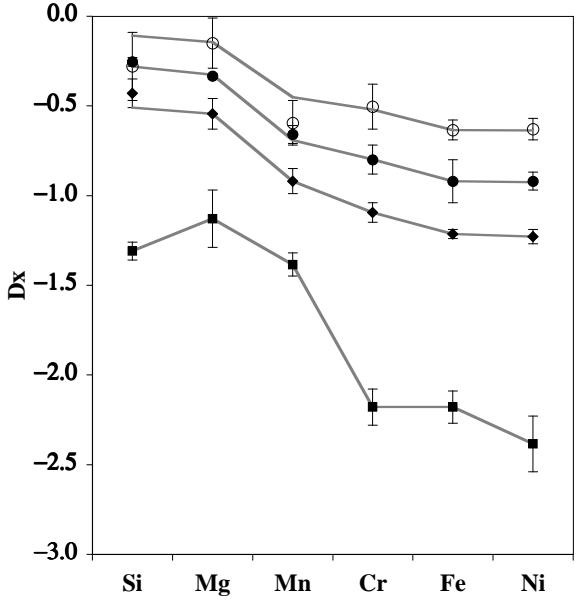}
\caption{
Comparison of observed and predicted 
interstellar depletion patterns in the Galaxy. 
Symbols: Galactic patterns observed in 
the \emph{cool disk} (squares), the \emph{warm disk} (diamonds), 
the \emph{warm disk plus halo} (filled circles), and 
the \emph{warm halo} (empty circles), 
  from Table 6 in SS96.  
Lines: model predictions obtained from the scaling law (\ref{ScalingLawISM}),
derived   
using the \emph{warm disk} as the  reference depletion
pattern and
normalized to fit the   observed
 Fe depletions
  using the values 
$r_1/r_2=1.058$, $r_3/r_2=0.935$, 
and $r_4/r_2=0.840$;
the parameters $\eta_\mathrm{x}$   are taken from Table 1 
($\eta_\mathrm{x,21}$). 
}
\label{Galaxy}
\end{figure}

Here we consider the   four types of depletion patterns
 discussed in SS96\footnote 
{
The Mg and Ni  column densities and   depletions
considered by SS96 have been updated using
revised values of the \ion{Mg}{2}
(Welty et al. 1999 and refs. therein) and
\ion{Ni}{2}   
(Fedchak \& Lawler 1999; Howk, Savage \& Fabian 1999)  oscillator strengths. 
}
(see Fig. \ref{Galaxy}).
In order of decreasing  
depletion, these patterns are representative of 
(1) cool clouds in the Galactic disk, 
(2) warm clouds in the disk, (3) disk plus halo clouds, 
and (4) warm halo clouds. 
The  lines of sight considered by SS96 also cover a significant range
of extinction properties, with strength of the 217.5 nm emission bump
as high as E(bump)=0.89 in HD\,24912 and as low as 
E(bump)=0.00 and 0.36 in HD 38666 and HD 116852, respectively
(Savage et al. 1985).  
We compared  pairs of the  4 SS96 depletion patterns to estimate
$\eta_{\mathrm{x},ij}$. The results are shown in Table 1, where the quoted
errors have been estimated propagating the uncertainties of
the fractions in dust. 
For some elements
the   $\eta_\mathrm{x}$ values obtained  
from different pairs of patterns show  
a different behavior
depending on whether the warm halo is considered or not. 
The $\eta_\mathrm{x}$ values obtained from  cool disk,
warm disk and warm disk-plus-halo patterns are in  agreement for
all elements considered.  
When we compare the warm halo   with the cool and warm disk,  the derived
parameters for Ni, Cr  and Mg
are still approximately constant\footnote{
Iron is not considered since    
$\eta_\mathrm{Fe} =0$ in all cases  by definition.
}.
However,  
  Mn and Si show different values, suggesting that for these elements the 
rate of incorporation into dust, or detachment from dust,     
 is different 
when we consider the transition to halo gas. 

The approximately constant values of the $\eta_\mathrm{x}$ parameters obtained 
for most elements  
over a wide range of interstellar conditions
  encourages us to use 
relation (\ref{ScalingLawISM}) for modeling all the Galactic depletion patterns.
In Fig. \ref{Galaxy}  we show  the results obtained using the
warm disk as the $i$-th reference pattern. 
One can see  that all the Galactic   patterns can be matched using
the set of values $\eta_\mathrm{x,21}$ listed in Table 1.
In spite of the lack  of constancy of
$\eta_\mathrm{Si}$ and $\eta_\mathrm{Mn}$ when
the warm halo pattern is considered,  the  model is capable of fitting 
also the Si and Mn
depletions in the warm halo within the observational errors. 
Therefore,   relation  (\ref{ScalingLawISM}) provides an empirical way to express
\emph{all} the 
different types of depletion patterns by varying only one parameter,
$r_j/r_i$. 
This is a significant improvement with respect to previous work, 
in which the fractions in dust were assumed to scale linearly with $r_j/r_i$
(i.e. $\eta_\mathrm{x}=0$ for all elements)
and in which by no means  was it possible  to model at the same type different   
depletion patterns.  

The fact that  all types of Galactic depletions 
can be modeled 
with  a single analytical expression 
 suggests  a continuity of the dust properties
in different types of interstellar clouds, 
consistent with results of previous work 
(Joseph et al. 1986; Joseph 1988; Sofia, Cardelli \& Savage 1994). 
  The existence of specific patterns of
depletions probably reflects the existence of well defined phases
of the ISM, each one with characteristic physical parameters which
determine the  properties of the dust, as proposed by Spitzer (1985). 
However,
in the transition from disk to halo gas important
changes in the dust properties must   occur to explain the sudden changes
of $\eta_\mathrm{x}$ values. These changes are quite significant for Si,
suggesting a  different behavior of this element 
as far as its ability to be incorporated in 
(or detached from) dust grains  
in different types of interstellar environments. 

\section{Estimate of the $\varepsilon_\mathrm{x}$ parameters from extragalactic
studies}

If the chemical composition of the medium is not constant,
it is necessary to  know the $\varepsilon_\mathrm{x}$ parameters
to apply  the scaling law (\ref{ScalingLawGeneral}). 
Among the possible choices of $\varepsilon_\mathrm{x}$, two values
have a special meaning: $\varepsilon_\mathrm{x}=0$ implies that the dust
composition is insensitive to variations of the overall composition;
$\varepsilon_\mathrm{x}=1$ implies that dust composition scales
in proportion to the composition of the medium, a possibility which
seems logical a priori. 
In addition, the possibility that $\varepsilon_\mathrm{x}=1$ is attractive as
this would imply that Galactic-type depletion patterns can be applied
to studies of galaxies with different chemical compositions.
%
%
In fact, when $\varepsilon_\mathrm{x}=1$, Eq. (\ref{ScalingLawGeneral})
becomes equal to Eq. (\ref{ScalingLawISM}), i.e. the depletion patterns
are of Galactic type independent of any change of the chemical
composition of the medium. 
As in this study we are not treating  the complex mechanisms of dust formation
and destruction, we  leave the possibility open that 
$\varepsilon_\mathrm{x}$ may take different values. 
We do not consider, however,   
  values $\varepsilon_\mathrm{x} < 0$.  
In fact, negative values  would imply that an  
increase of the   abundance of the element X in the medium
would lead to a decrease of its   abundance  in
the dust.

In order to  constrain  the parameters $\varepsilon_\mathrm{x}$
we must study   interstellar depletions in environments  with non-solar
abundances. These can be found, for instance, in the ISM of external galaxies.
To apply the scaling law to external galaxies we  assume
that   any   medium $i$
with specified $r=r_i$ and $a_\mathrm{x}=a_{\mathrm{x},i}$, 
even if  located in two arbitrary galaxies $A$ and $B$, is
characterized by a unique dust composition, i.e.  
$p_{\mathrm{x}}(r_i,a_{\mathrm{x},i})_{A}
=  p_{\mathrm{x}}(r_i,a_{\mathrm{x},i})_{B}$.
From this assumption, from (\ref{Fraction}) and from
(\ref{ScalingLawGeneral}) we   derive an expression which relates the 
depletions of the $j$-th interstellar medium in the
external galaxy $E$
to the  depletions of the $i$-th  medium in our Galaxy: 
\begin{equation}
\label{ScalingLawExtragal}
 f_{\mathrm{x},j}   
= \left(  {r_j \over r_i} \right)^{(1+\eta_\mathrm{x})}
\cdot 10^{\left( \varepsilon_\mathrm{x} - 1 \right) 
\left[ { {\rm X} \over {\rm Y} } \right]_{\mathrm{m},E} } 
\cdot  f_{\mathrm{x},i} 
~ ~,
\end{equation}
where 
\begin{equation}
\label{SquareBracketDef}
\left[ { {\rm X} \over {\rm Y} } \right]_{\mathrm{m},E}
\equiv \log \left( { {\rm X} \over {\rm Y} } \right)_{\mathrm{m},E}
- \log \left( { {\rm X} \over {\rm Y} } \right)_\sun  ~~,
\end{equation}
 $\left( { {\rm X} \over {\rm Y} } \right)_{\mathrm{m},E}$
is the  abundance in the medium of the external galaxy,
and $\left( { {\rm X} \over {\rm Y} } \right)_\sun$ is
the  abundance in the medium of our Galaxy, taken to be solar.
Eq. (\ref{ScalingLawExtragal}) is a generalization of Eq. (11)
  in Vladilo (1998), which can be derived from 
(\ref{ScalingLawExtragal}) for $\eta_\mathrm{x}=0$ and  $\varepsilon_\mathrm{x}=0$.

The scaling law (\ref{ScalingLawExtragal})   
can be used to estimate the depletions
in galaxies with non-solar abundances using 
a well-known Galactic  depletion pattern
as the reference pattern $i$.
As discussed in the previous section, 
Galactic studies indicate that the $\eta_\mathrm{x}$ parameters 
can be considered approximately constant under different types
of interstellar conditions, including variations of the 
physical state and of the
strength
of the 217.5 nm emission bump.  
This gives us confidence to use the values
derived in our Galaxy  for other galaxies as well,
even though, in principle, the adoption of Galactic
$\eta_\mathrm{x}$ values should be considered as
an a priori assumption. 
As far as the $\varepsilon_\mathrm{x}$ parameters are concerned,
we can probe their behavior   from
the   study of
nearby galaxies with non-solar abundances. 
For an external galaxy $E$,   sufficiently close
to measure  the gas-phase abundances, $( \mathrm{ X \over H} )_{\mathrm{g},E}$,
and to derive a set of reference  
abundances   from stellar studies, $( \mathrm{ X \over H} )_{\mathrm{m},E}$, the
depletions can be estimated from the expression
\begin{equation}
\label{ExternalDepletion}
D_{\mathrm{x},E} ~ = ~ 
\left[ \mathrm{ X \over H} \right]_{\mathrm{g},E} -
\left[ \mathrm{ X \over H} \right]_{\mathrm{m},E} 
~ = ~
\left[ \mathrm{ X \over H} \right]_{\mathrm{g},E} -
\left[ \mathrm{ Y \over H} \right]_{\mathrm{m},E} -
\left[ \mathrm{ X \over Y} \right]_{\mathrm{m},E} ~~, 
\end{equation}
where the terms in square brackets are defined as in Eq. (\ref{SquareBracketDef}).
It should be noted that not only the  model  depletions  
(\ref{ScalingLawExtragal}), 
but also the  measured depletions  
(\ref{ExternalDepletion})
 change according to the   abundance ratios
$[ \mathrm{ X \over Y} ]_{\mathrm{m},E}$. 
If we know 
these     ratios  from stellar studies 
we can compare the observed and predicted depletions  
and so  constrain the parameters $\varepsilon_\mathrm{x}$,
since we adopt $\eta_\mathrm{x}$ from Galactic studies. 
The Magellanic Clouds (MCs) are ideal targets for applying this
technique. A first application is presented in the next Section 4.1.

\subsection{Constraints from SMC interstellar studies}

Only a few  lines of sight have been observed at
high spectral resolution in the UV to study  interstellar depletions
in the MCs. We consider in this study  
the lines of sight to  the SMC stars Sk 78, Sk 108, and Sk 155,
for which  interstellar column densities of several metals have been obtained from
\emph{HST} observations 
(Welty et al. 1997, 2001; Koenigsberger et al. 2001) and 
   \ion{H}{1} column densities  
   from   {\em IUE} observations (Fitzpatrick 1985).

To derive the observed depletions  (\ref{ExternalDepletion})
we used the gas phase abundances $\left[ \mathrm{ X / H}
\right]_\mathrm{g,SMC}$  obtained
 from the total SMC interstellar column
densities  (Table 1 in Welty et al. 2001). 
%
%
For   the intrinsic SMC metallicity we adopted
$\left[ \mathrm{ Fe / H}
\right]_\mathrm{m,SMC} = -0.6 \pm 0.2$ dex from the typical values of  SMC stellar
abundances 
(Russell \& Dopita 1992; Welty et al. 1997). 
This range of metallicity  is  in agreement
with the SMC interstellar [Zn/H]  measurements in the same lines of sight,
taking into account that Zn is modestly depleted in the
Galactic ISM (Roth \& Blades 1995).\footnote
{
In the present paper we do not consider the detailed behavior of Zn 
 depletion  because this element is not
 included in the "standard" depletion patterns given by SS96. 
}  
To derive the model depletions  
(\ref{ScalingLawExtragal}), we used the fractions in dust $f_{\mathrm{x},i}$ of
the Galactic \emph{warm disk}   and the exponents $\eta_{\mathrm{x},21}$ that
  simultaneously fit
all the Galactic ISM patterns (Fig. \ref{Galaxy} and Table 1). 
All the model predictions are normalized to match the SMC iron depletions
by varying the  parameter $r_j/r_i$. 

As far as the intrinsic SMC abundances 
$\left[ \mathrm{ X / Fe} \right]_\mathrm{\mathrm{m},SMC}$ are concerned, we
first considered  the case of solar ratios and then a few non-solar
abundance patterns discussed below.
%
In all cases 
we obtain
$r_j/r_i=0.930, 0.802$, and $1.020$ for the sight lines   to
Sk 78, Sk 108, and Sk 155, respectively.
These values of $r_j/r_i$ are in the range 
typical of Galactic \emph{warm halo}, \emph{warm disk+halo},
and \emph{cold disk} clouds, respectively (compare with
the $r_j/r_i$ values in the caption to Fig. \ref{Galaxy}),
in agreement with the conclusions pointed out by
Welty et al. (1997, 2001).\footnote
{In addition, Welty et al. (2001) studied the multi-component
structure of the SMC absorptions toward Sk 155, while only the
total SMC column densities are considered here.}

\begin{figure} 
\plotone{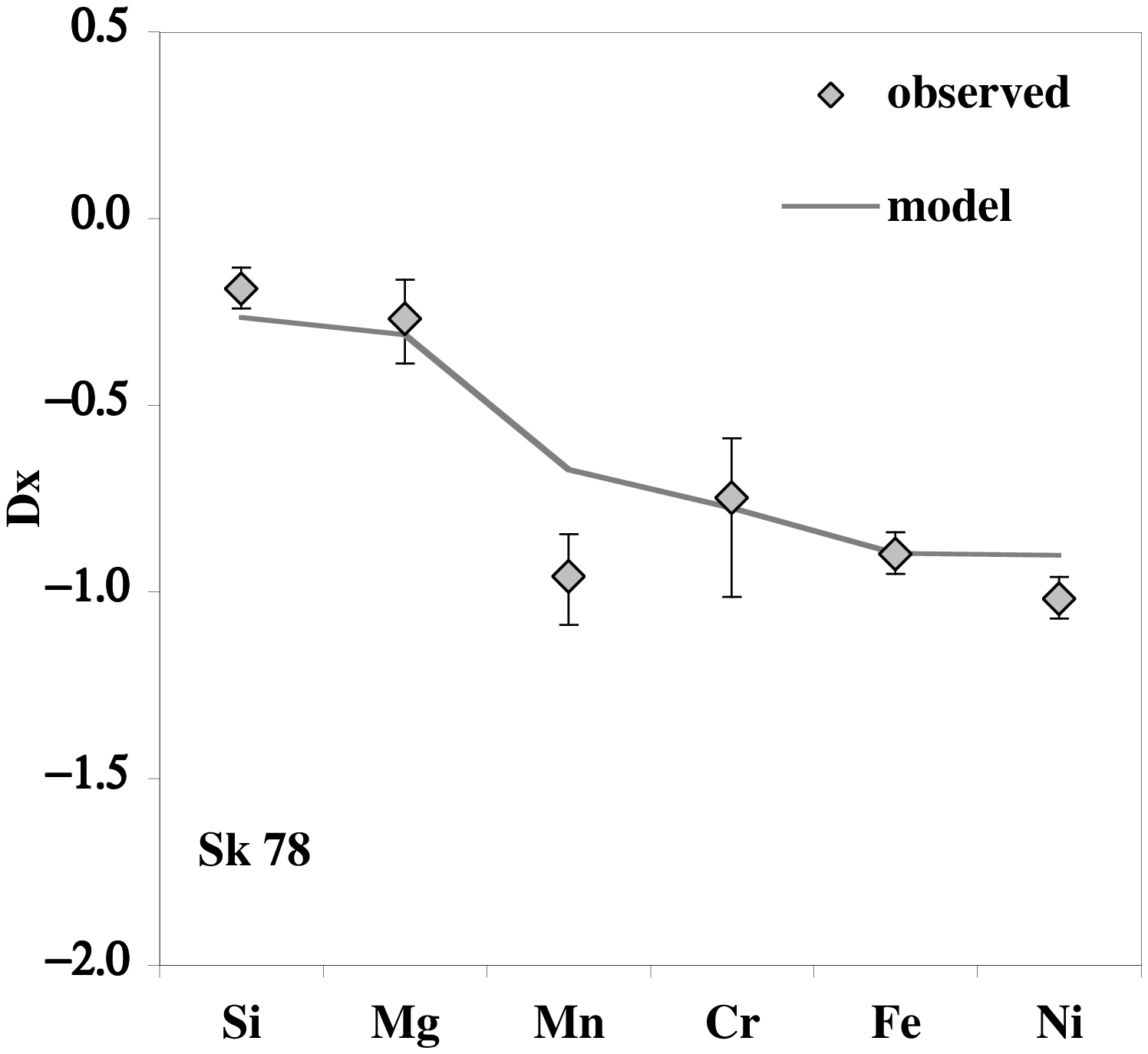} 
\caption{
Observed and predicted 
SMC depletion patterns in the  line of sight to Sk 78
derived from Eqs. (\ref{ExternalDepletion}) and 
(\ref{ScalingLawExtragal}), respectively,  for the case of
 solar  
$\left[ \mathrm{ X / Fe} \right]_\mathrm{ref,SMC}$ ratios. 
Symbols:  observed depletions obtained   
using the SMC column densities by Welty et al. (2001),
 $\log N$(\ion{H}{1})$_\mathrm{SMC}$ = $20.93 \pm 0.05$ (Fitzpatrick 1985),
and
$\left[ \mathrm{ Y / H} \right]_\mathrm{ref,SMC} = -0.6$ dex. 
Line: predicted  depletions obtained   using
the Galactic \emph{warm disk} as the reference pattern,
$r_j/r_i = 0.930$, and  
the parameters $\eta_\mathrm{x,21}$ (see Table 1 and Fig. \ref{Galaxy}).   
}
\label{figSk78solar}
\end{figure}

\begin{figure}
\plotone{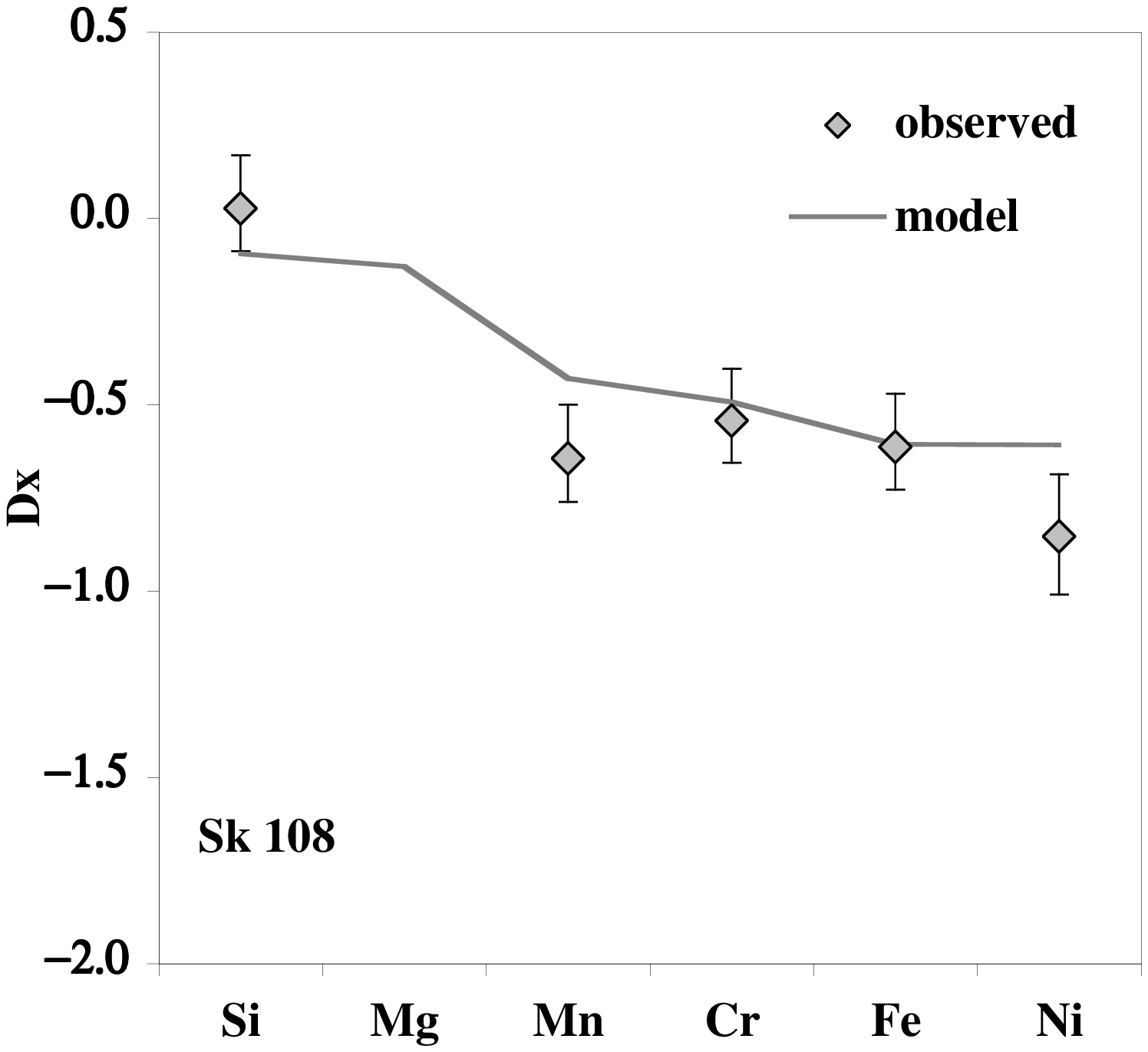} 
\caption{
Observed and predicted 
SMC depletion patterns in the  line of sight to Sk 108
derived from Eqs. (\ref{ExternalDepletion}) and 
(\ref{ScalingLawExtragal}), respectively,  for the case of
 solar  
$\left[ \mathrm{ X / Fe} \right]_\mathrm{ref,SMC}$ ratios.  
Symbols:  observed depletions obtained  
using the SMC column densities by Welty et al. (2001),
 $\log N$(\ion{H}{1})$_\mathrm{SMC}$ = $20.54^{+0.11}_{-0.14}$ (Fitzpatrick 1985),
and
$\left[ \mathrm{ Y / H} \right]_\mathrm{ref,SMC} = -0.6$ dex 
Line: predicted  depletions obtained   using
the Galactic \emph{warm disk} as the reference pattern,
$r_j/r_i = 0.802$, and  
the parameters 
$\eta_\mathrm{x,21}$ (see Table 1 and Fig. \ref{Galaxy}).
}
\label{figSk108solar}
\end{figure}

\begin{figure} 
\plotone{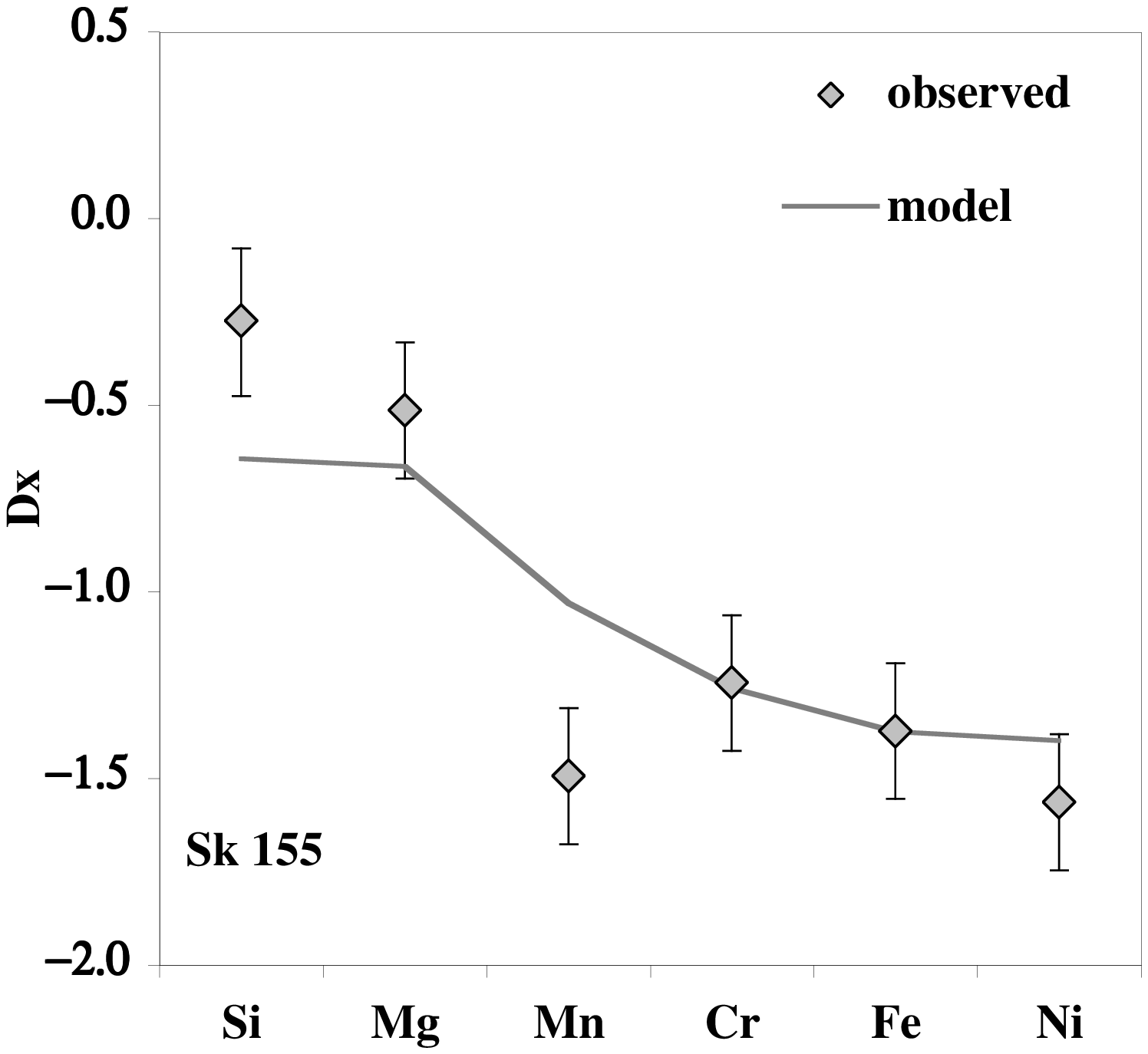} 
\caption{
Observed and predicted 
SMC depletion patterns in the  line of sight to Sk 155
derived from Eqs. (\ref{ExternalDepletion}) and 
(\ref{ScalingLawExtragal}), respectively,  for the case of
 solar  
$\left[ \mathrm{ X / Fe} \right]_\mathrm{ref,SMC}$ ratios. 
Symbols:  observed depletions obtained  
using the SMC column densities by Welty et al. (2001),
 $\log N$(\ion{H}{1})$_\mathrm{SMC}$ = $21.54 \pm 0.18$ (Fitzpatrick 1985),
and
$\left[ \mathrm{ Y / H} \right]_\mathrm{ref,SMC} = -0.6$ dex.
Line: predicted  depletions obtained  using
the Galactic \emph{warm disk} as the reference pattern,
$r_j/r_i = 1.020$, and  
the parameters 
$\eta_\mathrm{x,21}$ (see Table 1 and Fig. \ref{Galaxy}). 
}
\label{figSk155solar}
\end{figure}

For the case of  solar     ratios, i.e.  
 $\left[ \mathrm{ X / Fe} \right]_\mathrm{\mathrm{m},SMC}=0$, 
%
the results for the sight lines   to
Sk 78, Sk 108, and Sk 155
are shown in Figs. \ref{figSk78solar}, \ref{figSk108solar},
and \ref{figSk155solar}, respectively.
One can see that the  predicted depletions 
    do not   fit well the   observed ones. 
In particular, the Si and Mg measurements tend to lie above
the predictions  while the opposite is true for Mn 
and Ni. 
These disagreements  cannot be
attributed to the uncertainty of the metallicity level. 
In fact, 
similar results are  found adopting 
$\left[ \mathrm{ Fe / H} \right]_\mathrm{m,SMC} = -0.8 $ dex, 
and
$\left[ \mathrm{ Fe / H} \right]_\mathrm{m,SMC} = -0.4 $ dex, 
the two extreme   values of SMC metallicity still consistent with
 stellar data. 

The systematic deviations between observed and predicted depletions
  suggest that the SMC depletion patterns are peculiar
and/or that the SMC relative abundances are non solar.  
Welty et al. (2001) suggested that  
 the SMC dust may be essentially free of Si and Mg,
particularly
in the  line of sight  to  Sk 155  where they find
exceedingly high  [Si/Zn] and [Mg/Zn]
ratios. 
These authors made their analysis assuming solar SMC
  abundance ratios. 
In the present work we consider, in addition, the possibility that 
the SMC  abundances have modest deviations from
solar ratios.  
 

The SMC abundance pattern 
is rather uncertain. In Table 2  we list average     abundance 
ratios derived from SMC stellar studies, taken 
from   Russell \& Dopita (1992), Hill (1997), and Venn (1999).  
The observational spread is high and the measurements are often
consistent with solar values (Welty et al. 1997). However,
tentative evidence of departures from solar ratios is present in the data.
In fact, small departures  are expected at the metallicity of the SMC.
For instance, Milky Way stars with metallicities [Fe/H] $\simeq -0.6$ dex
do show characteristic non   solar ratios that
we also list in Table 2  
(from Ryan, Norris \& Beers 1996; RNB96). 
An enhancement of the $\alpha$/Fe ratios [Si/Fe] and [Mg/Fe]
and an underabundance of the [Mn/Fe] ratio
are  present   in metal-poor Galactic stars.  
Even if the SMC has probably undergone
a  chemical evolution different from that of the Milky Way, we 
still expect
to find a hint of
similar deviations in the SMC abundance pattern.
For instance, if the chemical evolution has proceeded with a lower rate of
(or a more episodic) star formation, the deviations 
of the $\alpha$/Fe ratios are probably less marked,
but still present (Matteucci 1991). 

\begin{figure} 
\plotone{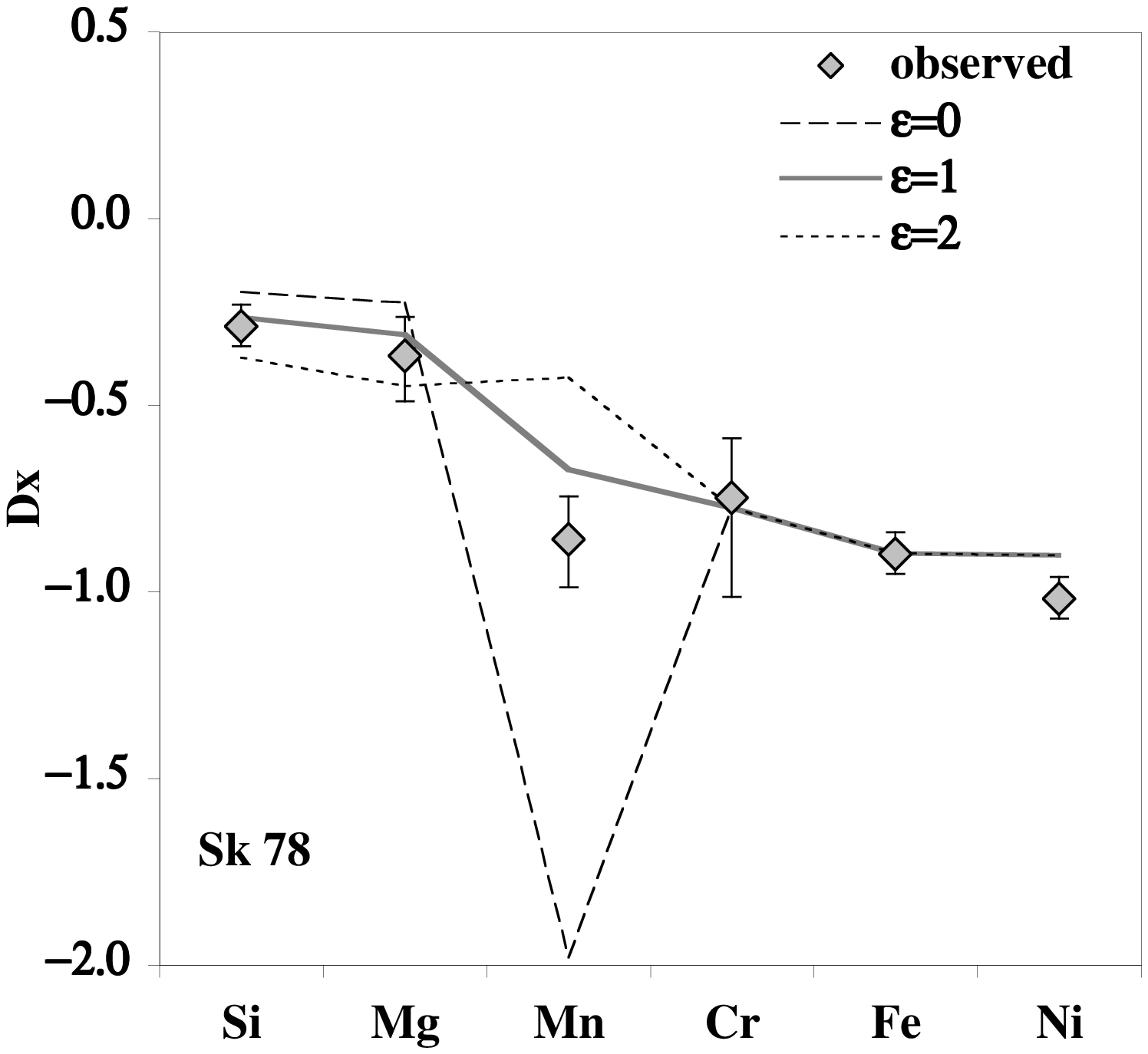} 
\caption{ 
Observed and predicted 
SMC depletions   toward Sk 78.
Same as in Fig. \ref{figSk78solar}, but with non solar 
$\left[ \mathrm{ X / Fe} \right]_\mathrm{ref,SMC}$ ratios
(abundance pattern "A" in Table 2).  
Lines:   depletions  predicted 
for different values of   $\varepsilon_\mathrm{x}$. 
Dashed line: $\varepsilon_\mathrm{x}=0$. 
Thick line: $\varepsilon_\mathrm{x}=1$. 
Dotted line: $\varepsilon_\mathrm{x}=2$. 
For $\varepsilon_\mathrm{Mn}=0$ all the atoms of Mn are predicted to be
in dust; a conventional
value $D_\mathrm{Mn,SMC}=-2$ dex is plotted in the figure. 
}
\label{figSk78A}
\end{figure}

\begin{figure} 
\plotone{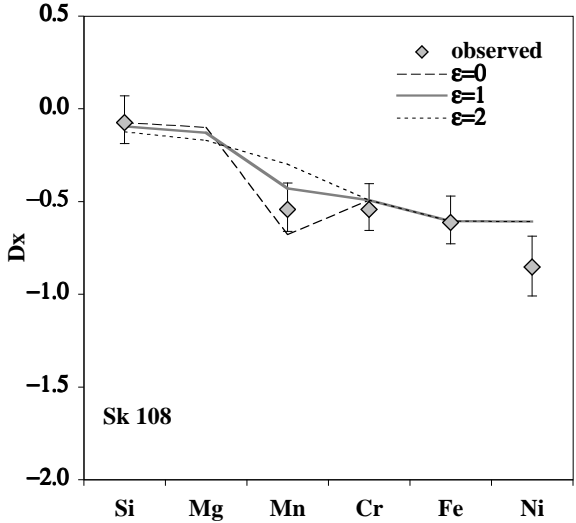} 
\caption{ 
Observed and predicted 
SMC depletions   toward Sk 108.
Same as in Fig. \ref{figSk108solar}, but with non solar 
$\left[ \mathrm{ X / Fe} \right]_\mathrm{ref,SMC}$ ratios
 (abundance pattern "A" in  Table 2).  
Lines:   depletions  predicted 
for different values of   $\varepsilon_\mathrm{x}$. 
Dashed line: $\varepsilon_\mathrm{x}=0$. 
Thick line: $\varepsilon_\mathrm{x}=1$. 
Dotted line: $\varepsilon_\mathrm{x}=2$. 
}
\label{figSk108A}
\end{figure}

\begin{figure} 
\plotone{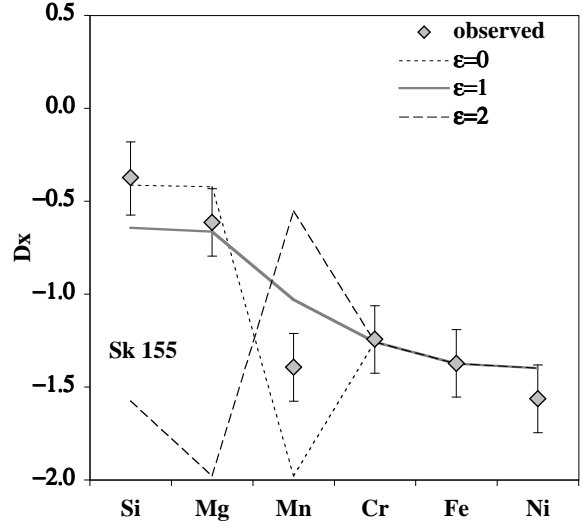} 
\caption{ 
Observed and predicted 
SMC depletions   toward Sk 155.
Same as in Fig. \ref{figSk155solar}, but with non solar 
$\left[ \mathrm{ X / Fe} \right]_\mathrm{ref,SMC}$ ratios
 (abundance pattern "A" in  Table 2).  
Lines:   depletions  predicted 
for different values of   $\varepsilon_\mathrm{x}$. 
Dashed line: $\varepsilon_\mathrm{x}=0$. 
Thick line: $\varepsilon_\mathrm{x}=1$. 
Dotted line: $\varepsilon_\mathrm{x}=2$. 
For $\varepsilon_\mathrm{Mn}=0$ and $\varepsilon_\mathrm{Mg}=2$
all the atoms of Mn and Mg are predicted to be
in dust;   conventional
values $D_\mathrm{Mn,SMC}=D_\mathrm{Mg,SMC}=-2$ dex are plotted in the figure.
}
\label{figSk155A}
\end{figure}

Since modest departures from solar  ratios  
are consistent with SMC stellar data
and are  expected from the comparison with Galactic stars, 
it is  worth examining if such departures can explain the
peculiarities of the SMC depletion patterns. 
To probe this possibility  we adopted the SMC   abundance pattern
 A  in Table 2  that  was chosen   
(1) to be consistent with the SMC stellar data and 
(2) to deviate from solar ratios in the
same sense of metal-poor Galactic stars, but  by a conservatively
low amount.  
The results obtained adopting the abundance pattern A to represent
$\left[ \mathrm{ Fe / Y} \right]_\mathrm{\mathrm{m},SMC}$ 
are shown in Figs. \ref{figSk78A}, \ref{figSk108A}, and \ref{figSk155A} 
for the sight lines   to
Sk 78, Sk 108, and Sk 155, respectively.
In each case,
three different values of  $\varepsilon_\mathrm{x}$ were considered,
namely $\varepsilon_\mathrm{x}=$  0, 1, and 2.  
For  $\varepsilon_\mathrm{x}=1$,  the
agreement between observed and predicted depletions is better than in the case
of intrinsic solar ratios.  For instance, the Si depletion 
is reproduced
within the errors   for Sk 78 and Sk 108. 
However   discrepancies are still present for other elements,
especially   in  the  sight-line to Sk 155. 
%
For $\varepsilon_\mathrm{x}$  
somewhat below unity the predictions are in 
   agreement with the observations.    
In fact, all the 
three depletion patterns, including the peculiar one
towards Sk 155, can be simultaneously fit with
  $0 < \varepsilon_\mathrm{x}   < 1$ for most elements considered.\footnote
{
The results for Ni depletions do not improve
with respect to the case of  
solar ratios because in
the abundance pattern  A  we adopt [Ni/Fe]=0. 
} 
For   $\varepsilon_\mathrm{x}=2$, the discrepancies are
even larger  than in the  case of solar ratios.

\begin{figure} 
\plotone{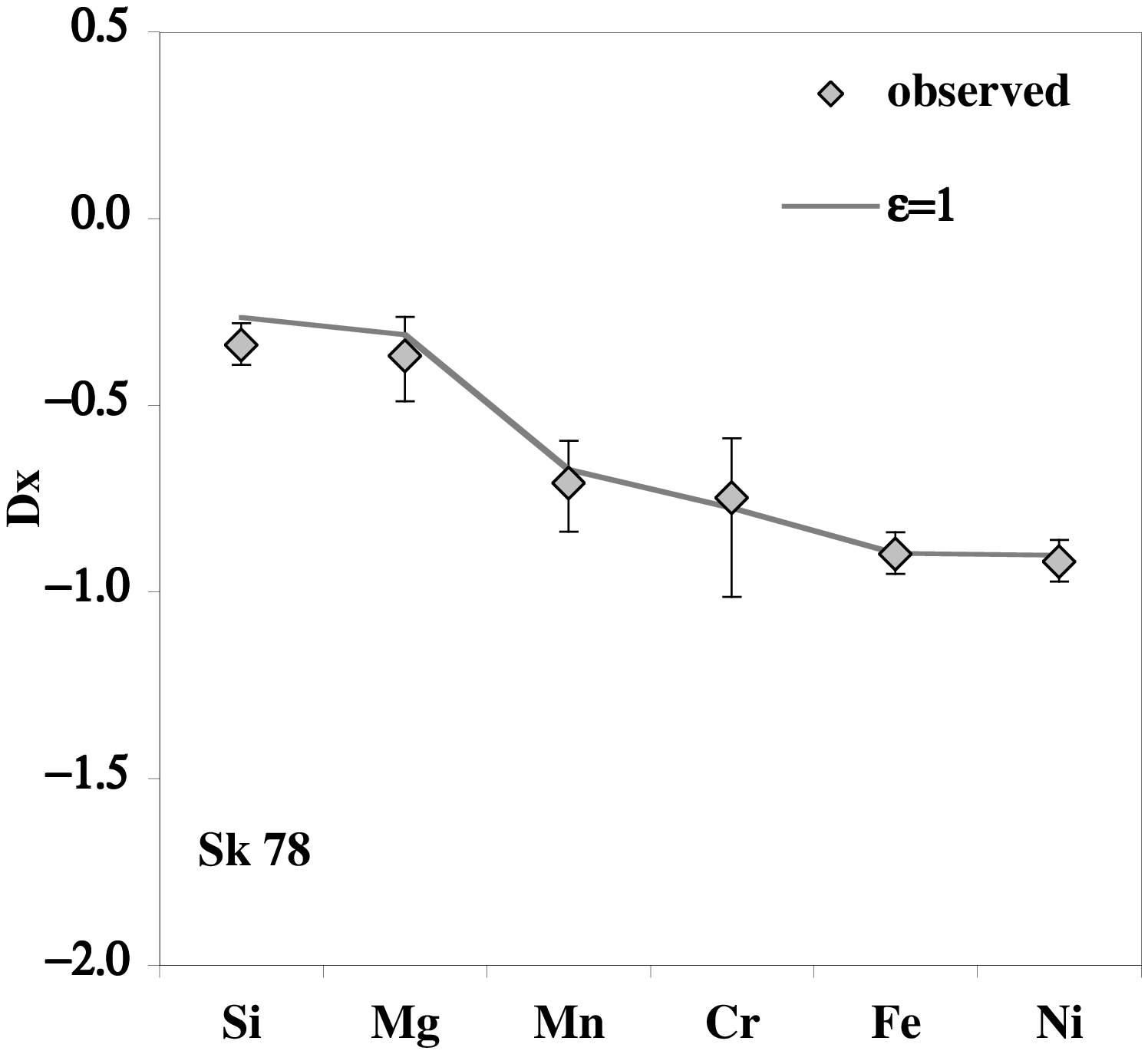} 
\caption{ 
Observed and predicted 
SMC depletions   toward Sk 78.
Same as in Fig. \ref{figSk78solar}, but with non solar 
$\left[ \mathrm{ X / Fe} \right]_\mathrm{ref,SMC}$ ratios
  (abundance pattern "B" in  Table 2).  
Thick line:   depletions  predicted for
 $\varepsilon_\mathrm{x}=1$. 
}
\label{figSk78B}
\end{figure}

\begin{figure} 
\plotone{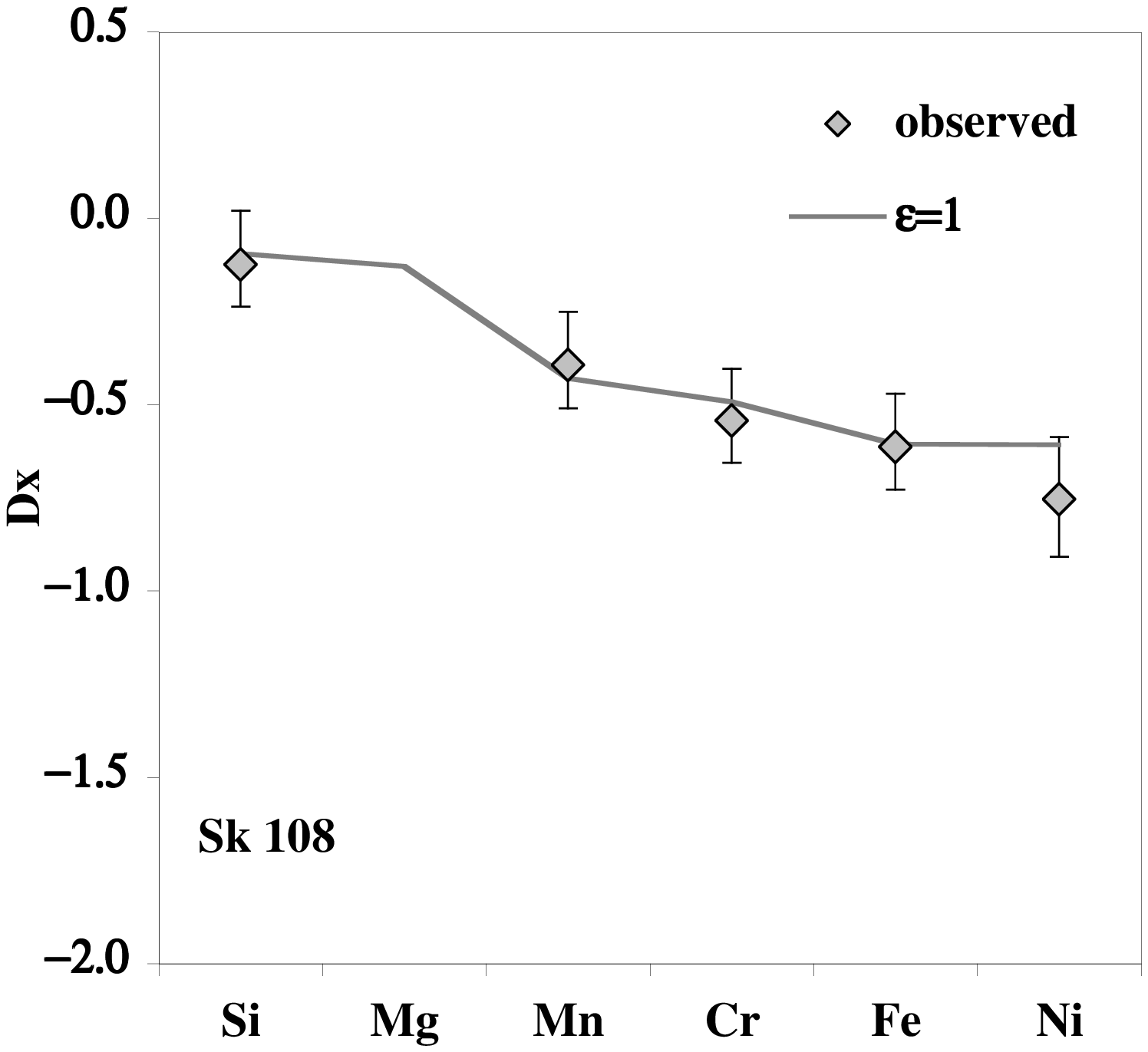} 
\caption{ 
Observed and predicted 
SMC depletions   toward Sk 108.
Same as in Fig. \ref{figSk108solar}, but with non solar 
$\left[ \mathrm{ X / Fe} \right]_\mathrm{ref,SMC}$ ratios
  (abundance pattern "B" in  Table 2).  
Thick line:   depletions  predicted for
 $\varepsilon_\mathrm{x}=1$. 
}
\label{figSk108B}
\end{figure}

\begin{figure} 
\plotone{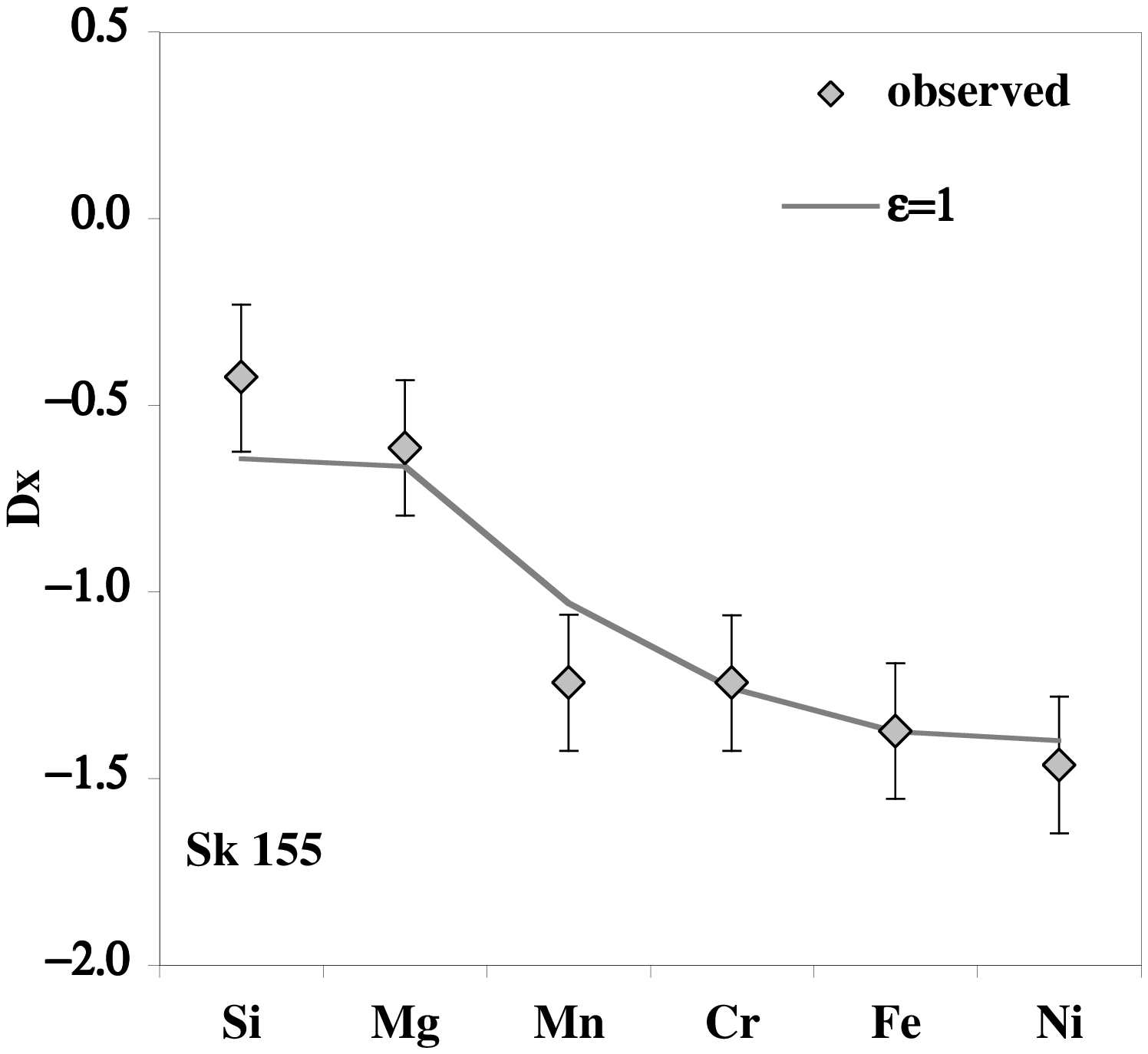} 
\caption{ 
Observed and predicted 
SMC depletions   toward Sk 155.
Same as in Fig. \ref{figSk155solar}, but with non solar 
$\left[ \mathrm{ X / Fe} \right]_\mathrm{ref,SMC}$ ratios
 (abundance pattern "B" in Table 2).  
Thick line:   depletions  predicted for
 $\varepsilon_\mathrm{x}=1$. 
}
\label{figSk155B}
\end{figure}

The depletions predicted for 
$\varepsilon_\mathrm{x} \neq 1$  tend to  jump 
to values much lower or higher than the observed  depletions. 
In some cases  $\varepsilon_\mathrm{x} \neq 1$ 
all the atoms of an element are predicted 
to be in dust (Figs. \ref{figSk78A} and \ref{figSk155A}).  
The curves obtained for $\varepsilon_\mathrm{x} = 1$
do not suffer from this inconvenience. 
In addition, the case  $\varepsilon_\mathrm{x}=1$ is attractive
for the reasons explained above (Section 4, first paragraph).
%
We therefore searched for a set of SMC relative abundances able to reproduce
the observed depletions   while keeping 
$\varepsilon_\mathrm{x}=1$ fixed for all   elements.  
By trial and error we find that the   abundance    pattern
B  in Table 2   fits all elements reasonably well.  
The results for the sightlines to Sk 78, Sk 108, and Sk 155
are shown
in Figs. \ref{figSk78B}, \ref{figSk108B}, and \ref{figSk155B}, 
 respectively. 
It is encouraging that a single set of abundances  
is able to  simultaneously fit the depletions observed
in all the three lines of sight
with Galactic type depletion patterns (i.e. with $\varepsilon_\mathrm{x}=1$).
 However, at the present time, some of the ratios
of the abundance pattern B seem  difficult
to reconcile  with SMC stellar data.  
 
In conclusion, it is not possible to model the SMC depletions
if the SMC abundances are strictly solar. 
We have two choices in modeling the observed depletions:
(1) modest departures from solar values (pattern A in Table 2)   
and $0 < \varepsilon_\mathrm{x} \leq 1$;
(2)   "Galactic type" depletions for all elements
 ($\varepsilon_\mathrm{x} = 1$),
but larger departures from solar ratios  
(pattern B in Table 2).  Values of $\varepsilon_\mathrm{x}$ in excess
of unity give depletion patterns inconsistent with the observations.

\section{Summary and conclusions}

An analytical expression has been derived that allows
dust depletions to be estimated in interstellar environments  
characterized by a wide range of chemical and physical properties. 
In practice, the depletions are estimated as a function of 
the dust-to-metals ratio, $r$, and of the relative abundance
of the element X in
the medium, $a_\mathrm{x}$,  by scaling a depletion pattern 
chosen as a reference.
The scaling law, given in Eq. (\ref{ScalingLawGeneral}), has been derived  
making no assumption  on the  mechanisms of dust formation,
accretion or destruction.  
No hypothesis has been made on the  extinction properties of the dust. 
The functional form (\ref{ScalingLawGeneral}) is always valid
for infinitesimal changes  $\delta r$ and  $\delta a_\mathrm{x}$.
For finite changes  $\Delta r$ and $\Delta a_\mathrm{x}$
the expression is valid
if the parameters  $\eta_\mathrm{x}$ and $\varepsilon_\mathrm{x}$
are constant. 
These parameters are essentially  derivatives of the 
relative abundance of the element X in the dust, $p_\mathrm{x}$.
In practice,
the parameters  $\eta_\mathrm{x}$ indicate how the chemical composition
of the dust
is affected by changes of the dust-to-metals ratio $r$. 
%
The parameters  $\varepsilon_\mathrm{x}$  
indicate  how the   composition of the dust is affected by changes
in the composition of the medium.


%
The scaling law can be applied to  interstellar clouds
of our Galaxy and of external galaxies with solar or non-solar
abundances --- including  Damped Ly $\alpha$ systems ---
 once the sets of parameters    $\eta_\mathrm{x}$
and $\varepsilon_\mathrm{x}$ are determined. 
The  assumption required to apply the scaling law to external galaxies 
is that   the dust chemical composition is a function of
$r$ and $a_\mathrm{x}$     valid in any type of interstellar medium.
This assumption is much more realistic than assumptions
previously adopted  in studies of DLA systems,
in which depletion patterns were estimated using   specific
types of Galactic dust.

The parameters  $\eta_\mathrm{x}$ and $\varepsilon_\mathrm{x}$
can be determined empirically by comparing observed depletion patterns
in interstellar media with different physical and chemical properties. 
We have used the typical Milky Way depletion patterns   to
estimate   the parameters $\eta_\mathrm{x}$ in 
environments with variable $r$ but constant (solar)    abundances.  
The resulting values of $\eta_\mathrm{x}$  are approximately constant for
most elements in a 
wide range of physical conditions, supporting the validity of 
the scaling law (\ref{ScalingLawGeneral}),
 at least as far as  the dependence on $r$
is concerned.
With the derived set of $\eta_\mathrm{x}$ parameters,
the scaling law is able to simultaneously fit 
all the typical  Milky Way depletion patterns
  by only varying $r$.

The parameters $\varepsilon_\mathrm{x}$ can be determined empirically
by studying interstellar
depletions   of galaxies with known abundances
different from the solar ones. We have applied this technique
to the SMC, for which the intrinsic abundances are constrained from stellar
studies and depletion data are available for three lines of sight. 
The observed SMC depletions can be modeled with the scaling law
if the  SMC abundance ratios are slightly non solar.
The required departures from solar ratios
are consistent with the results of SMC stellar studies
and with expectations based on Galactic stellar abundances
of stars having the same metallicity of the SMC.  
The agreement between observed
and predicted depletions is obtained at values 
$\varepsilon_\mathrm{x} \simeq 1$ or somewhat below unity. 
With this choice of parameters  
the scaling law is able to simultaneously fit  the
three available SMC depletion patterns,
including the anomalous pattern observed toward Sk 155.   
Values $\varepsilon_\mathrm{x} > 1$ are excluded by the present analysis. 
The uncertainty of the SMC stellar abundances prevent  
firmer conclusions or   more constrained $\varepsilon_\mathrm{x}$ values.

The capability of the scaling relation presented here
to match Galactic and SMC depletion patterns is rather encouraging.
To probe the general validity of the scaling law 
it is  desirable to measure   depletions 
in a larger number of  extragalactic lines of sight.
However, these studies will yield stringent constraints only if 
the intrinsic abundances   of the external galaxies, mostly based
on stellar observations, will  be determined with accuracy.  
A major step in this direction can be
made in the next few years by observing 
 a relatively large
number of stars of the Magellanic Clouds at high spectral resolution. 
These time-consuming efforts  will be rewarded  by
the possibility of modeling interstellar depletions and deriving
dust-corrected abundances in high redshift galaxies 
observed as QSO absorption systems. 
An implementation of the scaling relation for deriving dust-corrected
abundances in Damped Ly\,$\alpha$ systems will be presented in a separate paper.

\acknowledgements
 I wish to thank Daniel Welty an anonymous referee for
 suggestions that have improved the present work.

\clearpage

\begin{deluxetable}{lccccc} 
\tablecaption{Empirical determination of the $\eta_\mathrm{x}$ parameters\tablenotemark{a}}
\tablewidth{0pt}
\tablehead{ 
\colhead{X} &
\colhead{$\eta_\mathrm{x,21}$} & \colhead{$\eta_\mathrm{x,31}$} & 
\colhead{$\eta_\mathrm{x,32}$} & \colhead{$\eta_\mathrm{x,42}$} & \colhead{$\eta_\mathrm{x,41}$}
}
\startdata 
Si & $+4.68 \pm 1.77$\tablenotemark{b}	&
$+5.27\pm 0.60$&$+5.78\pm 1.89$\tablenotemark{b}  &
$+0.87\pm 2.53$\tablenotemark{b} & $+1.70\pm 1.94$ 		\\      
Mg & 
$+3.60\pm 1.50$ &	$+3.48 \pm 0.35$ &	$+3.37 \pm 1.29$  &
$+3.46 \pm 3.98$ & $+3.49 \pm 3.10$ 		\\   
Mn & $+0.53 \pm 0.41$ &
$+0.69 \pm 0.27$ &	$+0.82 \pm 0.60$  & $-0.18 \pm 0.51$ &
$-0.02 \pm 0.39$ 		\\ 
Cr & $+0.37 \pm 0.20$ &	$+0.37 \pm 0.29$ &
$+0.36 \pm 0.56$  & $+0.45 \pm 0.66$ & $+0.43 \pm 0.52$ 		\\
Fe & $+0.00 \pm 0.07$ & $+0.00 \pm 0.32$ & $+0.00 \pm 0.59$ 
& $+0.00 \pm 0.19$ & $+0.00 \pm 0.15$ 		\\
Ni	&  $+0.01 \pm 0.11$	& $+0.02 \pm 0.13$	& $+0.03 \pm 0.26$ 
& $+0.03 \pm 0.21$ & $+0.02 \pm 0.17$ 		\\  
\enddata  

\tablenotetext{a}{
Each parameter  $\eta_\mathrm{x}$ is derived 
from relation (\ref{ScalingLawISM}) by comparing the fractions in dust
of the corresponding element in two different types of Galactic depletion patterns.
The fraction in dust is estimated from  relation (\ref{fraction2}). 
The two indices of each parameter indicate which pair of
 depletion patterns is considered. The indices correspond to:
(1) cool disk, (2) warm disk, (3) warm disk+halo, (4) warm halo.
The parameters are normalized
 in such a way that  $\eta_\mathrm{Fe}=0$
for any pair of depletion patterns considered. 
The parameters $\eta_\mathrm{x,34}$ are not estimated because the  
depletion patterns (3) and (4) are almost overlapping within the errors. 
}
\tablenotetext{b}{  
$D_\mathrm{Si} = -0.51$ dex  adopted for the warm disk depletion of Si.
This choice of $D_\mathrm{Si}$, which lies in the range given by SS96,  
gives a better simultaneous fit of all the depletion patterns.}

\end{deluxetable}

\clearpage

\begin{deluxetable}{lcccccc} 
\tablecaption{Abundance ratios in metal-poor Galatic stars and in the SMC}
\tablewidth{0pt}
\tablehead{ 
\colhead{ } &
\colhead{RNB96\tablenotemark{a}} & 
\colhead{RD92\tablenotemark{b}} & \colhead{H97\tablenotemark{c}}& \colhead{V99\tablenotemark{d}} &
\colhead{A} & \colhead{B}   
}
\startdata 
$\left[\mathrm{Si/Fe}\right]$ ~~~~~&    +0.19 & +0.15 $\pm$ 0.22  & $-0.14 \pm 0.19$ & $+0.26 \pm 0.19$ & +0.10 & +0.15 \\    
$\left[\mathrm{Mg/Fe}\right]$ ~~~~~&    +0.12 & +0.07 $\pm$ 0.18  & $+0.16 \pm 0.22$ & $-0.01 \pm 0.12$ & +0.10 & +0.10 \\  
$\left[\mathrm{Mn/Fe}\right]$ ~~~~~& $-$0.30  & +0.17 $\pm$ 0.34  & \nodata          & \nodata          & $-0.10$ & $-0.25$ \\
$\left[\mathrm{Cr/Fe}\right]$ ~~~~~& $-$0.02  & +0.09 $\pm$ 0.24  & $-0.31 \pm 0.18$ & $-0.02 \pm 0.13$ & +0.00 & +0.00 \\
$\left[\mathrm{Ni/Fe}\right]$	~~~~~&    +0.00 & +0.27 $\pm$ 0.24  & $-0.38 \pm 0.14$ & \nodata          & +0.00 & $-0.10$ \\ 
 \enddata  

\tablenotetext{a}{
Abundances of metal-poor Milky Way stars 
at metallicity [Fe/H] $\simeq -0.6$ dex. Values taken in
correspondence of the "midmean vector" defined by Ryan et al. (1996). 
}
\tablenotetext{b}{
From Table 1  by Russell \& Dopita (1992), corrected when necessary
for the meteroritic solar abundances by Anders \& Grevesse (1989).  
The abundances are relative to Galactic 
stars of same spectral type. 
}
\tablenotetext{c}{
Average values of 6 K-type supergiants in the SMC
(Hill 1997; Table 3).  Mg measured only in 3 stars. 
}
\tablenotetext{d}{
Average values of  10 A-type supergiants in the SMC  
(Venn 1999; Table 6). Only measurements of singly ionized   species have been considered. 
The abundances are relative to Galactic A-type supergiants.  
}
\tablecomments{
Solar ratios taken from the meteoritic
values by Anders \& Grevesse (1989). 
The  errors have been computed propagating the errors of each of the two elements 
considered in the ratio.
}
\end{deluxetable} 

\clearpage

\clearpage

\clearpage

\end{document}